\documentclass[fleqn,usenatbib]{mnras}

\usepackage{newtxtext,newtxmath}
\usepackage[T1]{fontenc}
\usepackage{ae,aecompl}
\usepackage{url}
\usepackage{graphicx}
\usepackage{xspace}
\usepackage{amsfonts}
\usepackage{pifont}
\usepackage{textcomp}
\usepackage{float}
\usepackage{subfig}
\usepackage{multirow}
\usepackage{adjustbox}

\title[Circular Polarimetry of mCVs]{Circular Polarimetry of Suspect Wind-accreting Magnetic pre-Polars}

\author[P. Hakala et al.]{
Pasi Hakala$^{1}$\thanks{E-mail: pahakala@utu.fi}, Steven G. Parsons$^{2}$, Thomas R. Marsh$^{3}$, Boris T. G\"ansicke$^{3}$, Gavin Ramsay$^{4}$,
\newauthor 
Axel Schwope$^{5}$, J.J. Hermes$^{6}$
\\
$^{1}$Finnish Centre for Astronomy with ESO (FINCA), Quantum, University of Turku, FI-20014, Finland\\
$^{2}$Department of Physics and Astronomy, University of Sheffield, Sheffield, S3 7RH, UK\\
$^{3}$Department of Physics, University of Warwick, Coventry CV4 7AL, UK\\
$^{4}$Armagh Observatory and Planetarium, College Hill, Armagh, BT61 9DG, UK\\
$^{5}$ Leibniz Institute for Astrophysics Potsdam (AIP), An der Sternwarte 16, 14482 Potsdam, Germany\\
$^{6}$Department of Astronomy, Boston University, 725 Commonwealth Ave., Boston, MA 02215, USA\\
}

\begin{document}
\label{firstpage}
\pagerange{\pageref{firstpage}--\pageref{lastpage}}

\outer\def\gtae {$\buildrel {\lower3pt\hbox{$>$}} \over 
{\lower2pt\hbox{$\sim$}} $}
\outer\def\ltae {$\buildrel {\lower3pt\hbox{$<$}} \over 
{\lower2pt\hbox{$\sim$}} $}
\newcommand{\Msun}{$M_{\odot}$}
\newcommand{\lsun}{$L_{\odot}$}
\newcommand{\Rsun}{$R_{\odot}$}
\newcommand{\solar}{${\odot}$}
\newcommand{\kep}{\sl Kepler}
\newcommand{\ktwo}{\sl K2}
\newcommand{\tess}{\sl TESS}
\newcommand{\swift}{\it Swift}
\newcommand{\Porb}{P_{\rm orb}}
\newcommand{\nuorb}{\nu_{\rm orb}}
\newcommand{\eplus}{\epsilon_+}
\newcommand{\eminus}{\epsilon_-}
\newcommand{\cd}{{\rm\ c\ d^{-1}}}
\newcommand{\MdotL}{\dot M_{\rm L1}}
\newcommand{\Mdot}{$\dot M$}
\newcommand{\Mdsolar}{\dot{M_{\odot}} yr$^{-1}$}
\newcommand{\Ldisk}{L_{\rm disk}}
\newcommand{\src}{KIC 9202990}
\newcommand{\ergscm} {erg s$^{-1}$ cm$^{-2}$}
\newcommand{\rchi}{$\chi^{2}_{\nu}$}
\newcommand{\chisq}{$\chi^{2}$}
\newcommand{\pcmsq} {cm$^{-2}$}

\maketitle

\begin{abstract}
We present results from a circular polarimetric survey of candidate detached magnetic white dwarf -- M dwarf binaries obtained using the Nordic Optical Telescope, La Palma. We obtained phase resolved spectropolarimetry and imaging polarimetry of
seven systems, five of which show clearly variable circular polarisation. The data indicate that these targets have white dwarfs with magnetic field strengths $>80$\,MG. Our study reveals that cyclotron emission can dominate the optical luminosity at wavelengths corresponding to the cyclotron emission harmonics, even in systems where the white dwarfs are only wind-accreting. This implies that a very significant fraction of the the stellar wind of the companion star is captured by the magnetic white dwarf reducing the magnetic braking in pre-CVs. Furthermore, the polarimetric confirmation of several detached, wind-accreting magnetic systems provides observational constraints on the models of magnetic CV evolution and white dwarf magnetic field generation. We also find that the white dwarf magnetic field configuration in at least two of these systems appears to be very complex.
\end{abstract}


\section{Introduction}
Cataclysmic variables (CVs) are semi-detached binaries where a cool donor, usually a main-sequence star (i.e. the secondary) fills its Roche-lobe and matter is transferred via Lagrangian L$_{1}$ point overflow and accreted by the primary white dwarf (WD).
In the case of non-magnetic WDs, an accretion disc is formed, but if the WD magnetic field is strong enough ($\sim1$\,MG or higher)
the disc is at least partly disrupted and the inner accretion flow onto the surface of the magnetic WD (MWD) follows
the magnetic field lines. In case of strong magnetic fields (i.e. $\sim10$\,MG or more), the entire accretion disc is disrupted and the
ballistic accretion stream leaving the L$_{1}$ point will be eventually re-directed along the magnetic field lines of the MWD (within its Alf\'en radius) and impacts the MWD near the magnetic pole(s) (see \citealt{Cropper1990} for an early review). Another consequence of the strong magnetic field of the WD is the synchronicity of the MWD spin period with the orbital period of the system: these synchronous systems 
are called polars (or AM\,Her binaries).

In addition to affecting the accretion geometry of these systems, the magnetic fields of both the WD and the secondary star play a vital role in the binary evolution. These sources initially form as a consequence of common envelope (CE) phase of binary evolution \citep{Pacz1967} as one of the binary components is swallowed by the outer envelope of the other, evolved, binary component. This leads to a rapid loss of angular momentum and thus the binary orbit spirals in. Finally the red giant phase ends, and as a result, a binary consisting of a main sequence star and a compact object emerges. These systems have periods from hours up to several days. The orbit of this binary shrinks further as angular momentum is lost due to both magnetic braking and gravitational radiation (see \citealt{Pacz1967} and \citealt{VZ1981}). At some point the low mass main-sequence star will start to fill its binary Roche lobe and, as the orbital shrinkage continues, mass will start to overflow the Lagrangian L$_{1}$ point and accrete on the primary component giving rise to a CV in the case of a WD or an X-ray Binary for a neutron star or black hole. 

However, in Low Accretion Rate Polars (LARP, \citealt{2002ASPC..261..102S}) the secondary star does not quite fill its Roche lobe, but there is still some indication of matter being accreted on the WD as beamed cyclotron emission is observed on/above the WD surface in the optical light curves. 
The hot spots are indicative of a strong magnetic field of the WD, that channels the accretion onto its magnetic poles. As there is no Roche lobe overflow, the implication is that the accretion must take place via stellar wind from the secondary star that is captured by the WD magnetosphere. These systems have also later been called PREPs (i.e. pre-polars, \citealt{2009A&A...500..867S}) following the realisation that most of them do not contain a Roche lobe filling donor. %However, the LARP term might be more appropriate as it does %not imply any particular evolutionary history, only %wind-accretion.

\begin{table*}
  \begin{tabular}{lcccccc}
    \hline
Source & Date & Filter/grism & Mode & UT range & Exp. time (sec) & Notes\\
    \hline
2MASS J01294349+671530$^1$ &	16.10.2020 &	Grism 5 &		circ.sp.pol. &       20:46-06:31 &  180 & (three 20-30min gaps) \\

SDSS J222918.95+185340.2$^2$ &		17.10.2020 &	Grism 5 & 		circ.sp.pol. &       19:37-19:46 & 200 & (4 circ. pol. spectra) \\

				&	17.10.2020 &	$R_{\mathrm{Bess}}$ &		circ.im.pol. &       19:52- 23:14 & 30 & \\
	
				&	16.11.2020 &        $R_{\mathrm{Bess}}$ &		circ.im.pol. &        20:43-22:35 & 20 & \\	

SDSS J030856.55$-$005450.7$^3$ &  		18.10.2020 &	Grism 5 &		circ.sp.pol. &       00:24-00:45 & 300 &  (4 circ. pol. spectra, very faint) \\ 	

    			&		18.10.2020 &	$R_{\mathrm{Bess}}$ &	    circ.im.pol. &        00:50-05:05 & 30 & (some 60sec due to bad weather) \\

ZTF J014635.73+491443.1$^4$ &	    11.12.2020 &	Grism 10 &		circ.sp.pol. &        23:21-01:48 & 500 & \\ 

				&	15.12.2020 &	Grism 10 &		circ.sp.pol. &        23:48-02:25 & 180 & \\

SDSS J075015.11+494333.2$^2$ &	04.03.2021 &	Grism 5 &		circ.sp.pol. &        20:59-01:24 & 300 & \\ 

SDSS J085336.03+072033.5$^2$ &	04.03.2021 &	$R_{\mathrm{Bess}}$ &		circ.im.pol. &    20:03-00:05 & 30 & \\

%SDSS J1220+5654 & 05.03.2021 &  $R_{Bess}$ &		circ.im.pol. &    00:13-01:55 & 30 & \\	

SDSS J121209.31+013627.7$^5$ & 08.03.2021 &      $i$ &	circ.im.pol. &      01:42-02:34 & 45 & (cirrus) \\

\hline
  \end{tabular}
  \caption{The Observing log where we indicate the source name; the start date of the observations; the filter in the case of imaging polarimetry or Grism in the case of spectroscopic polarimetry; the UT time range of the observations; the exposure time of each image and any notes on the observations.($^1$\citet{Krush2020},$^2$\citet{2021MNRAS.502.4305P},$^3$\citet{Becker2011},$^4$\citet{Guidry2021},$^5$\citet{2005ApJ...630L.173S}}
  \label{table1}
\end{table*}
Recently \citet{2021NatAs...5..648S} performed detailed computations of magnetic CV evolution, where they suggested that the stronger magnetic field strengths observed in polars, as compared to single WDs, could be due to WD magnetic field generation by a dynamo process \citep{2017ApJ...836L..28I}, as the crystallized WD is spun up by accreted matter. They managed to explain the observed properties of magnetic CV population (and the lower field strengths observed in single WDs) remarkably well. In their evolutionary model the detached pre-polars are in fact a product of a \emph{ non-magnetic} CV phase, preceding the polar phase. These systems are predicted to populate certain orbital period intervals and thus building up the observed population is crucial for testing the model. 
This has also possible implications for the magnetic braking theory of the binary period evolution \citep{Li1994}. If a large fraction of the stellar wind is {\sl not} carried out of the system, the magnetic braking is at least greatly reduced, slowing down the evolution of the binary towards shorter orbital periods and potentially making the observed population of CVs much older.  

For this study, we have compiled a sample of candidate "pre-polars", where the secondary star does not fill its Roche lobe. The selected sources all have previous observed properties suggesting the presence of possible cyclotron emission (i.e. evidence for cyclotron humps in the optical spectrum; extraordinary colours in light curves or excess emission in some passband). The individual sources are introduced and discussed separately in the subsections below. The cyclotron emission spectrum
is best identified from continuum circular polarisation. Circular spectropolarimetry allows 
reliable identification of cyclotron harmonic emission and therefore determination of the magnetic field strength. High time resolution imaging circular polarimetry, on the other hand, can be used to map the location(s), size(s), shape(s) and magnetic polarity of the cyclotron emission region(s) in detail.

We have carried out circular polarimetric observations at the 2.56\,m Nordic Optical Telescope (NOT) in order to detect and measure the
strength of cyclotron emission from these systems. We now proceed to describe the observations, data analysis and results from our campaign. 

\begin{figure*}
  \begin{center}
\hspace{0mm}  
\vspace{0mm}
  \includegraphics[width=0.74\textwidth]{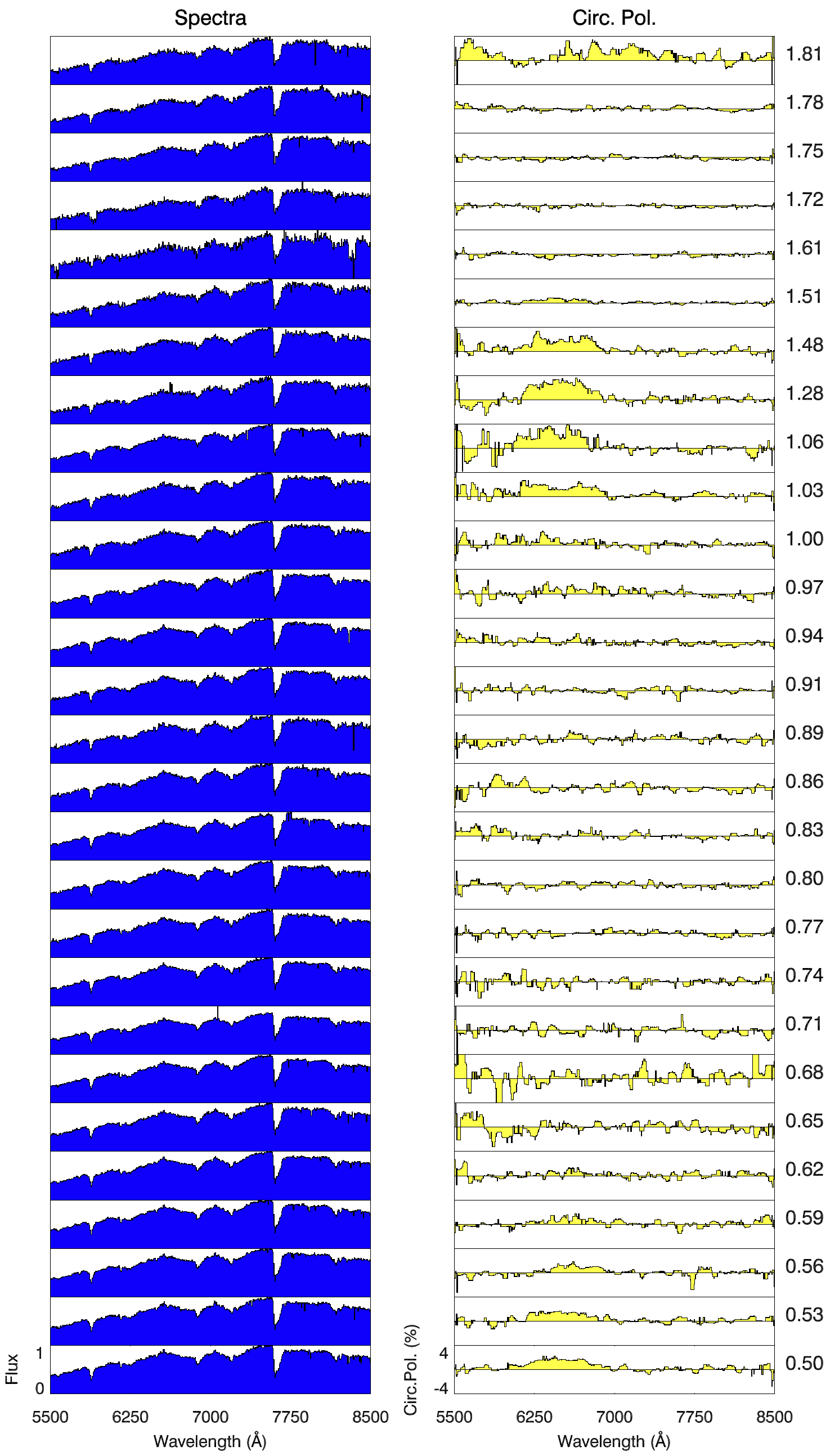}    
\vspace{0mm}
  \caption{Circular spectropolarimetry of 2MASS\,J0129+6715 over 1.3 orbital cycles (left flux, right circ. polarisation). The corresponding orbital phases calculated from the photometric ephemeris of \citet{Krush2020} are shown. The polarisation vanishes later during the first cycle than during the second (e.g. compare orbital phases 0.533$-$0.562 vs. 1.511).}
    \label{figure1}
    \end{center}
\end{figure*}

\section{Observations}

The NOT observations were mainly carried out during two observing runs in October 2020 and March 2021 with some additional `fill-in' data taken in December 2020 (see Table \ref{table1} for the observing log). The observations were obtained
using ALFOSC, a multimode imager/spectrograph capable of both polarimetric imaging and spectropolarimetry. Brighter systems were observed in circular spectropolarimetric mode, whilst fainter systems were subject to imaging circular polarimetry in 
single passband ($R$ band) selected to maximize the detection probability for a typical field strength. The details of different modes are also listed in Table 1. ALFOSC was equipped with the e2V back illuminated deep depletion 2048x2064 CCD with 15 $\mu$m pixel size resulting in a full FOV of 6.4x6.4'.
However, only a partial window mode, together with on-chip binning, was used in this study. The spectra
were taken using either grism \#5 or \#10 (see table 1.). These grisms provide spectral coverages of
5000-10700\,\AA\ and 3300-11000\,\AA\ respectively. The grisms have formal resolutions of $R=415$ and $R=105$
using a 1 arcsec slit. However, in our case, we used a 1.8" polarimetric slitlet and additionally binned the CCD by a factor of two in both directions. Thus our effective resolution is of the order of $R\sim 50-200$ at maximum. The polarimetric observing modes involve using a calcite block together with a rotatable $\lambda$/4 plate, rotated with steps of 90$^{\circ}$. Obtaining data at four different angles of the retarder plate allows for automatic cancelling of transmission effects in the optical system, both for spectropolarimetic and polarimetric imaging data. However, using only two angles
(0$^{\circ}$ and 90$^{\circ}$) is also possible and provides better time resolution for variable 
sources.

All the data were bias-subtracted and the spectra were flatfielded using the halogen lamp exposures. The circular polarisation imaging time series data were obtained under autoguiding with a quarterwave-plate rotating at 90$^{\circ}$ steps. Flatfielding is not required for this mode.
We checked for systematic circular polarisation by observing a bright nearby field star both in imaging and in spectropolarimetric modes. This yielded instrumental circular polarisation of 0.248$\pm$0.064\%. We also combined 49 spectra of SDSS J0750+4943 observed in the spectropolarimetric mode and measured the circular polarisation
within the spectral region 6500-9000\AA\ that does not show any cyclotron emission. We found the instrumental polarisation to be 0.052$\pm$0.016\%. Compared to the observed changes of several, or even dozens of, per cent present in our targets, the level of instrumental polarisation is insignificant. Furthermore,
during imaging polarimetry observations, the FOV contained field stars as a check for both circular
polarisation zero level and for photometric comparison.
We used IDL scripts for both the data reduction and optimal extraction of spectra (OPTSPECEXTR package\footnote{J. Harrington, {\url{https://physics.ucf.edu/~jh/ast/software/optspecextr-0.3.1/}}}). 
Imaging aperture photometry was carried out using our own scripts based on IDL astronomy library.

\section{Analysis \& discussion of individual sources}

In this section we proceed to describe each target and their related datasets separately.
We provide a brief introduction for each target, followed by the description of the data as well as the relevant
modelling and results. We then give an overview of the results and discuss them jointly in the following section. 

\subsection{{\it 2MASS\,J01294349+671530}}

2MASS J012943+6715 is a very recently discovered eclipsing 7.15\,h period pre-CV, where the secondary does not quite fill its Roche lobe \citep{Krush2020}. The eclipses of the WD show a very unusual excess in $R$ band when compared to the other bands. It was suggested by \citet{Krush2020} that this could be explained as a result of cyclotron emission from the magnetic WD accreting from a stellar wind  with a cyclotron hump centered at/near the $R$ band. 

We have obtained a set of circular spectropolarimetry data of the system covering 8.75\,h (with a 2.5\,h gap during the run). The spectra cover a wavelength range from 5500 to 8500\,\AA\. This
produced 31 circular polarisation spectra which are shown in Fig \ref{figure1}. Unfortunately the 2.5\,h gap in the data meant that altogether only about 0.6 of 
the orbital phase was covered (twice). There is a clear cyclotron emission hump centered at
around 6400-6500\,\AA\, verifying \citet{Krush2020}'s explanation for the unusual brightness of the WD in the R-band. No other humps are visible in the 5500-10000\,\AA\ wavelength range. This suggests that the hump is of 
low harmonic order. If we identify the hump as the 2nd harmonic of the cyclotron base frequency then the implied magnetic field strength is 81-85\,MG (see eg. \citet{WL2000} for magnetic field determination based on cyclotron harmonics). This would imply the base frequency cyclotron hump at around 12900\,\AA. However, if we choose
the 3rd harmonic instead, the magnetic field would be 53-57\,MG and the 2nd harmonic would be centered at 9740\,\AA. As the spectrum 
becomes quite noisy towards the red end, we cannot exclude the existence of the second harmonic in the spectra, but given these
data, we prefer the 81-85\,MG field interpretation. 

As our data overlaps a fraction of the orbital period (0.6), we can also examine the level of synchronicity of the WD spin period with the binary orbital period.
The individual circularly polarised spectra show at least some evidence for the cyclotron hump until orbital phase 0.59
during the first orbit. However, during the second orbital cycle, the cyclotron emission disappears between orbital phases
0.48-0.51 (Fig \ref{figure1}). There is thus some evidence that the WD spin period could be $\sim$10\% shorter than the orbital period. It is likely though, that the asynchronicity would
have been detected by \citet{Krush2020}, even if they do not have any polarimetric data. There are seven asynchronous polars known \citep{Rea2017}, most of which have the WD spin and orbital periods within 1-2\,\%. However, there are three systems i.e. RX J0838.7-2827 \citep{Rea2017}, Paloma \citep{Schwarz+07, Joshi2016} and IGR J19552+0044 \citep{Bernardini2013}, where the WD spin period is of the order of 80-90\% of the orbital period.  However, none of these have an orbital period comparable to 2MASS\,J0129+6715 which would make it very unusual amongst the magnetic CVs if it too was highly asynchronous. Further polarimetry is required to resolve this possibility.

It is interesting to note that it is not immediately clear how a system like 2MASS\,J0129+6715 could be produced by the evolutionary model of \citet{2021NatAs...5..648S}. While the WD appears cool enough to have started crystallization and a brief period of asynchronous rotation is expected in this model, the problem is that at the long period of this system, 7.15h, the donor stars in CVs are far less over-inflated than in shorter period systems. Therefore, when the magnetic field emerges from the WD and detaches the system, these longer period systems will very rapidly come back into contact, since the donor star will not substantially shrink within it's Roche lobe. To reach the measured Roche lobe filling factor of only 0.86 would be very difficult to achieve at this orbital period (see supplementary figure 1 in \citealt{2021NatAs...5..648S} for example). 

\subsection{{\it ZTF J014635.73+491443.1}}

\begin{figure}
  \begin{center}
  \vspace{0mm}
  \includegraphics[width=0.5\textwidth,angle=0]{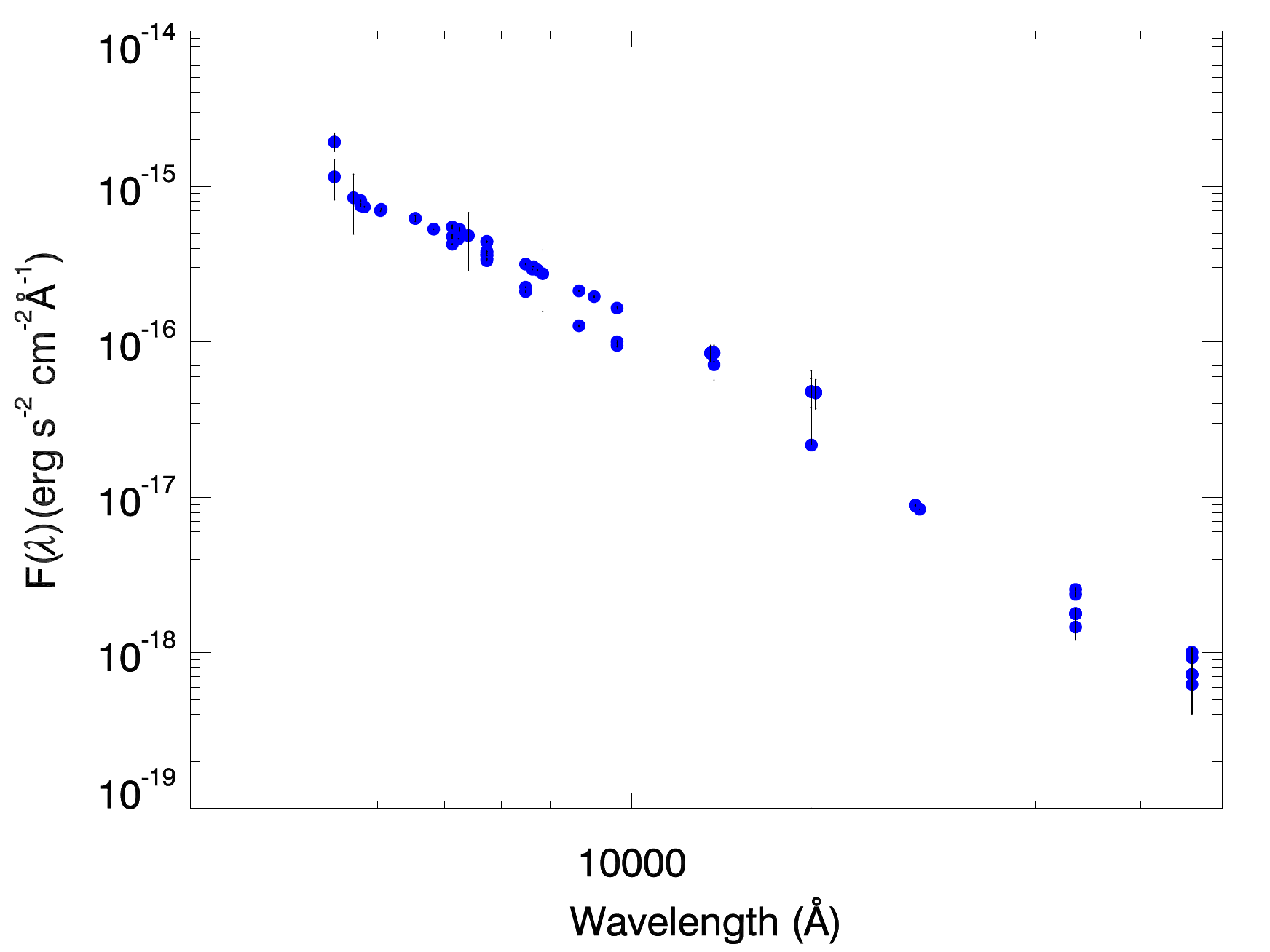}    
\vspace{-3mm}
  \caption{The SED of ZTF J0146+4914 from the Vizier catalogue (doi.org/10.26093/cds/vizier). There is no indication of any companion star.}
%  doi.org/10.26093/cds/vizier
    \label{figuresed}
    \end{center}
\end{figure}

\begin{figure*}
  \begin{center}
\vspace{-10mm}
  \includegraphics[width=0.9\textwidth]{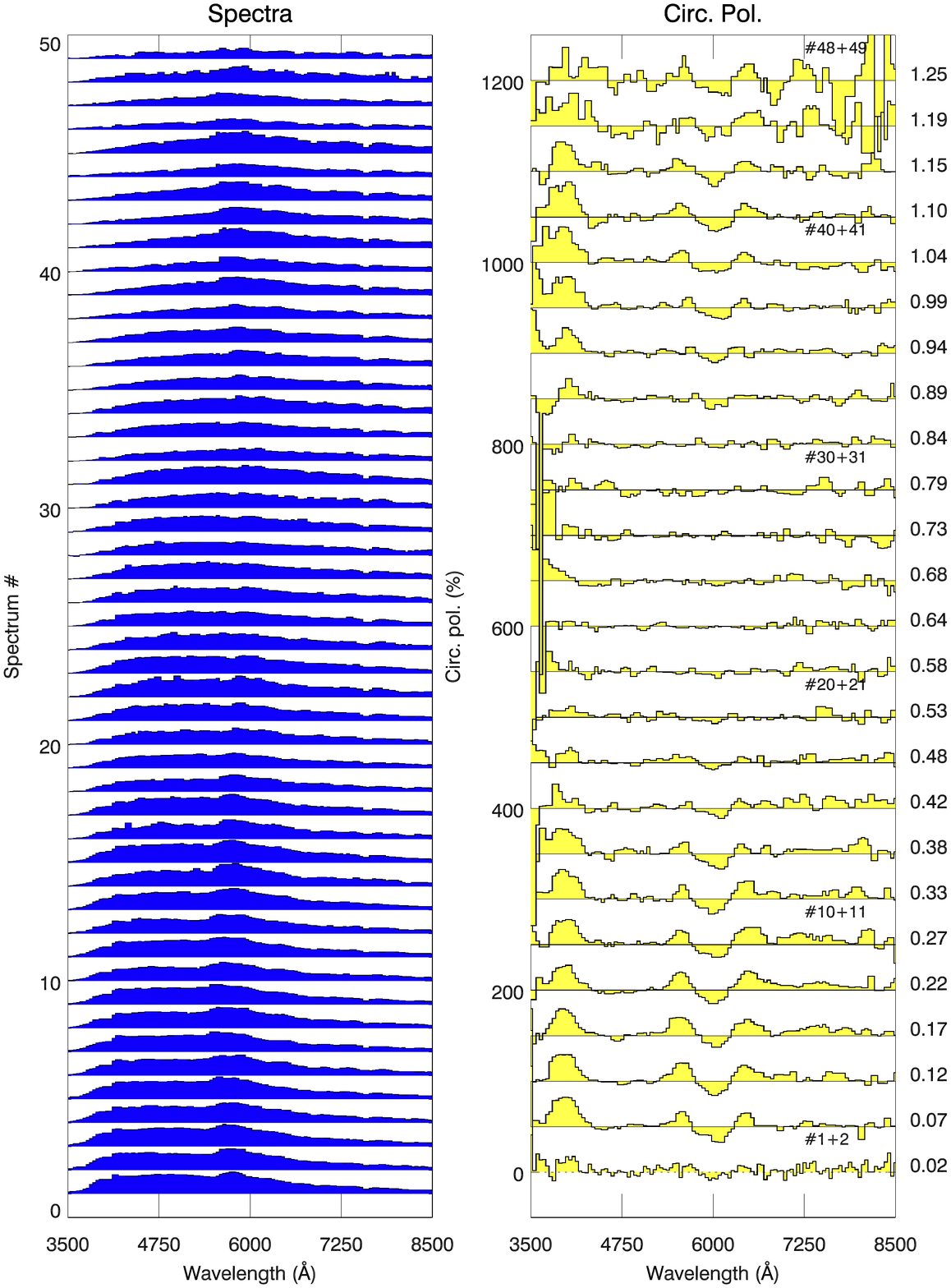}    
\vspace{0mm}
  \caption{Circular spectropolarimetry of ZTF J0146+4914. The left column shows the flux spectra vs time, offset in Y-direction. The right column shows the corresponding circular polarisation. The temporal separation of the flux spectra is half of that of polarimetry. The labels on the right side of some polarimetric spectra indicate the orbital phase (arbitrary 0.0 phase) coverage and the numbers inside the panel the relevant flux spectra they are based on. The orbital modulation of cyclotron humps is clearly visible both in the flux spectra and the circular polarisation spectra.}
    \label{figure2}
    \end{center}
\end{figure*}

\begin{figure}
  \begin{center}
\hspace{-10mm}  
\vspace{5mm}
  \includegraphics[width=0.53\textwidth]{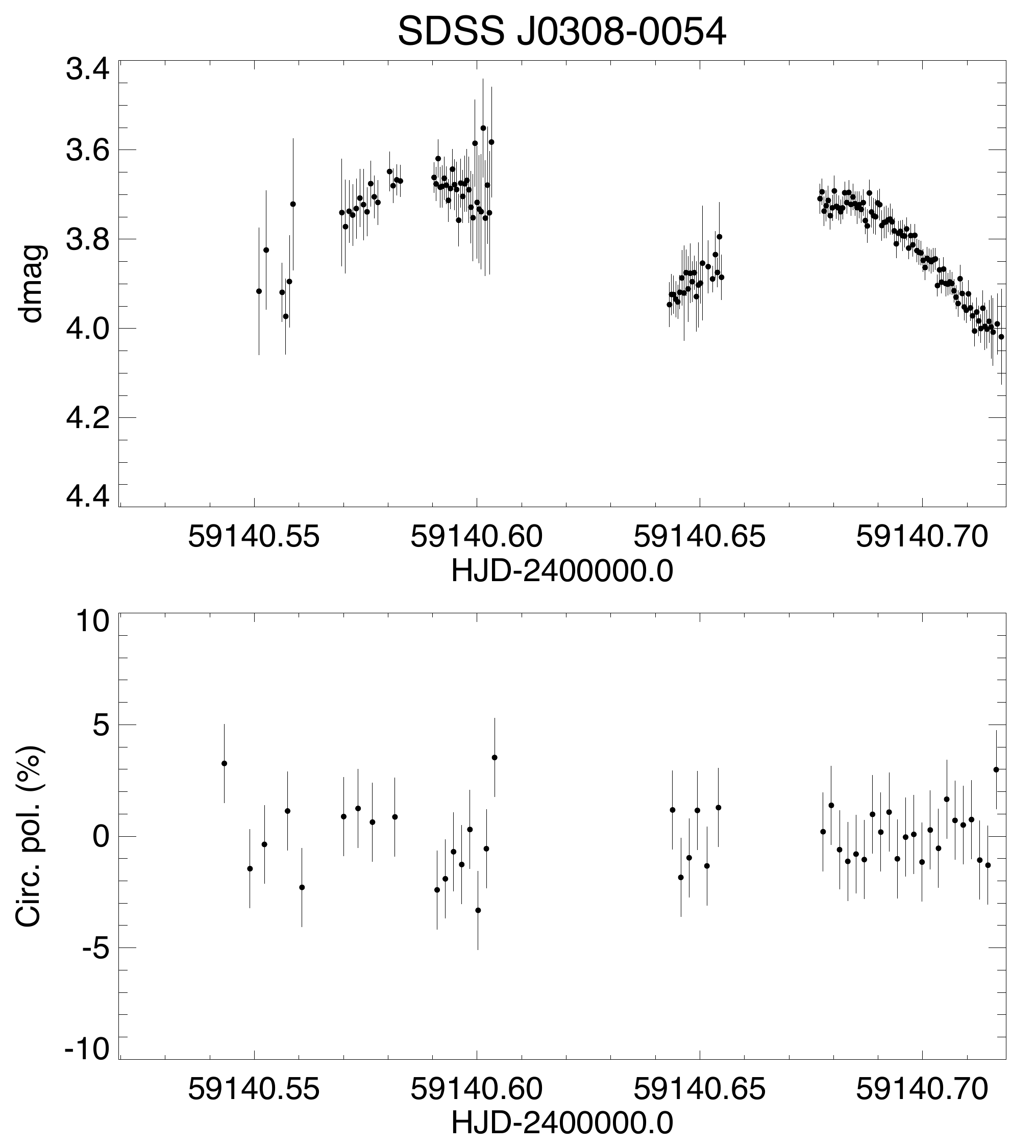}    
\vspace{0mm}
  \caption{Circular $R$ band imaging photopolarimetry of SDSS J0308-0054. Top panel: the differential magnitude compared to a nearby comparison star. Lower panel: the circular polarised light curve.}
    \label{figure3}
    \end{center}
\end{figure}
ZTF J0146+4914 was discovered as a candidate polar with a photometric period of 2.057 h by \citet{Guidry2021}. 
The system shows cyclotron hump features in the optical spectra, compatible with a WD magnetic field of $\sim$89\,MG.
The distance of (56.3 $\pm$ 0.3\,pc, \citet{B-J2018}) could potentially establish ZTF J0146+4914 as the closest
known polar. However, there are no clear emission lines visible in the spectra, suggesting a wind accreting system.
Furthermore, there is also no clear sign of the secondary in the SED Fig \ref{figuresed}, indicating a very low mass companion, probably a brown dwarf. This, together with the orbital period, makes the system a very likely period bouncer (i.e. a system that has passed the period minimum in its orbital evolution and is now evolving towards longer periods) and thus a very old system. 

We have attempted to characterise the donor brown dwarf by extracting the WISE \citep{2010AJ....140.1868W} magnitudes from the CatWISE \citep{2020ApJS..247...69E} catalogue and comparing them against
the theoretical DA WD WISE magnitudes obtained by interpolating from the Montreal WD model grid \citep{2006AJ....132.1221H,2011ApJ...730..128T} \footnote{\url{http://www.astro.umontreal.ca/~bergeron/CoolingModels}} using $T=8700$K and $\log(g)=8.355$ from \citet{2021MNRAS.508.3877G}. We detect a very small IR excess in the WISE bands W1 and W2. The corresponding  absolute WISE W2 magnitude of 13.65 would imply a $\sim$T8 donor based on Fig.\,29 of \citet{2011ApJS..197...19K}. However, the colour of the IR excess, $W1-W2=0.96$, suggests a much earlier donor of T2-T3 spectral class (Fig.\,1 of \citealt{2011ApJS..197...19K}). It is possible that the system contains a late T type donor, that has been slightly heated up by its WD companion to match these figures.

In fact, ZTF J0146+4914 appears to belong to a distinct subpopulation of systems, which include so far EF Eri \citep{1979Natur.281...48W}, SDSS J1212+0136 \citep{2005ApJ...630L.173S}, IL Leo \citep{2007ApJ...654..521S,2021MNRAS.502.4305P} and the 
SDSS J1250+1549 \& SDSS J1514+0744 \citep{2012MNRAS.423.1437B}. All of these share the following properties: their donor stars are very likely brown dwarfs, they show prolonged states of low accretion and they have orbital periods below the period gap ($\la$2hrs). Furthermore, all of them also exhibit WD temperatures of the order of 10\,000\,K (ZTF J0146+4914 appears in the eDR3 WD catalogue of \citealt{2021MNRAS.508.3877G} with a WD effective temperature of 8700K). The magnetic field strengths of these systems vary from 13 up to 89\,MG; effectively covering the full range typically observed in polars. These systems may represent highly evolved polars, i.e.  the equivalent of WZ\,Sge dwarf novae among the non-magnetic CVs. The fact that one of these system (ZTF J0146+4914) is located only 56\,pc away suggests that these systems are likely to be fairly common, as was already suggested by \citet{2011MNRAS.411.2695P} and \citet{2012MNRAS.423.1437B} based on their space density estimates.

The previously detected cyclotron humps in the optical spectra of ZTF J0146+4914 are centered at $\sim$4000\,\AA\ and $\sim$6000\,\AA\ and show somewhat different behaviour as a function of orbital phase. Especially the hump at 6000\,\AA, assumed to be the 2nd harmonic of the cyclotron base frequency, changes remarkably in both width and shape over the orbital period.

We obtained circular spectropolarimetry of ZTF J0146+4914 at two epochs in December 2020 using ALFOSC with grism \#10 covering a wavelength range of 3500-8500\AA. We show the higher time resolution spectropolarimetric time series in Fig \ref{figure2}. There are a couple of striking features in the spectra. Firstly, we detect the changes in width and shape of the 6000\,\AA\ cyclotron hump over the orbital period, discovered by \citet{Guidry2021}. Secondly, we note that the sign of the circular polarisation is reversed both at the red and blue ends of the cyclotron hump. This reversal persists for the duration the cyclotron region is visible over the orbital period, suggesting that there are in fact two emission regions visible at the same longitude on the WD surface. These would have different polarity, but similar visibility over the orbital period. It also appears that the cyclotron hump produced by the positive region is wider than the one produced by the negative region, leading to the observed sign reversal effect. Since the humps are centered
at the same wavelength, the magnetic field strengths at the two accretion regions are very similar. The different hump widths are thus likely to originate from differences in optical depth due to plasma density/temperatures together with the different viewing angles (however, see our modelling attempts on source SDSS J0750+4943 in \S 3.4, which shows very similar behaviour). We are not aware that such wavelength dependent cyclotron hump sign reversal has been reported before in any magnetic CV. For instance, there is no suggestion of this in the spectropolarimetry of LARPs in \citet{2005ApJ...630.1037S} or \citet{2007ApJ...654..521S}.  

\subsection{{\it SDSS J030856.55-005450.7}}

This source is an eclipsing dM4+WD binary with a 4.5\,h period \citep{Becker2011} . There is some tentative evidence that the eclipse depth may be variable (from CCD photometry with 0.5m class telescopes), which might indicate the presence of changing accretion rate/cyclotron emission on the WD surface. 
We first attempted circular spectropolarimetry, but the source was too faint to obtain useful data. The source was then observed for 4.2\,h
in imaging circular polarisation mode (Fig \ref{figure3}) under variable transparency conditions. The $R$ band photometry reveals the orbital
modulation, but there is no indication of any circular polarisation.

\subsection{{\it SDSS J075015.11+494333.2}}

\begin{figure*}
  \begin{center}
  \hspace{-17mm}
  \includegraphics[width=1.1\textwidth]{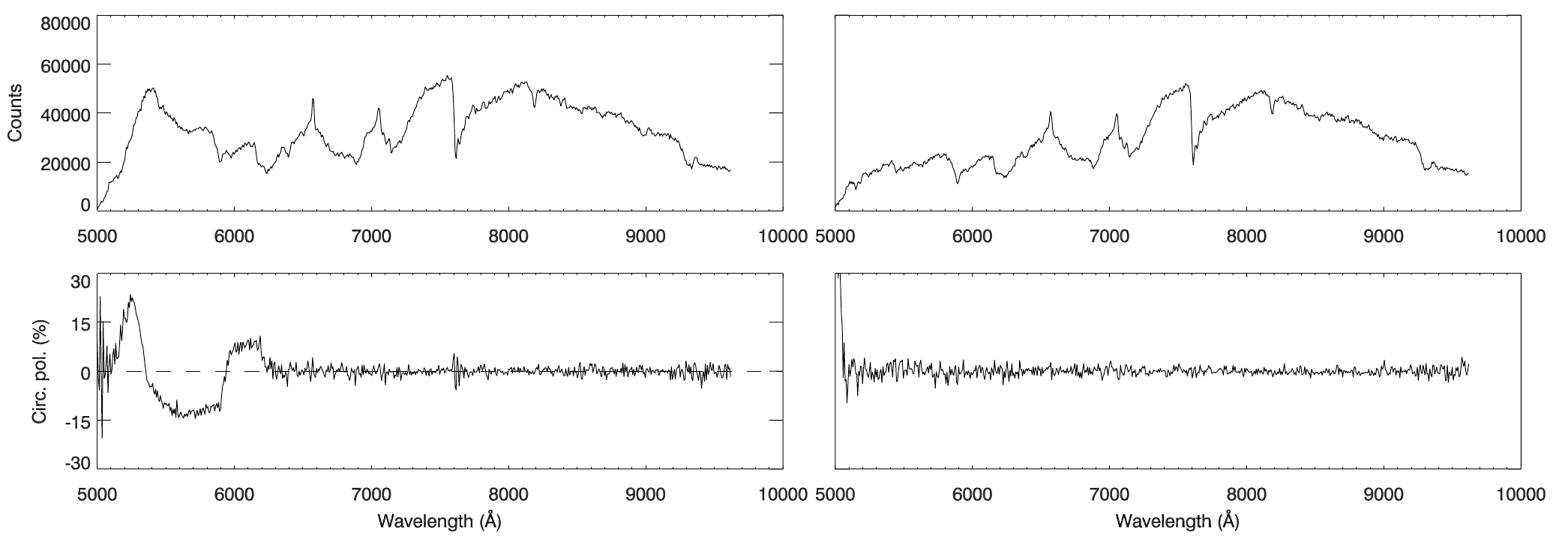} \vspace{10mm} 
  \includegraphics[width=1.05\textwidth,angle=0]{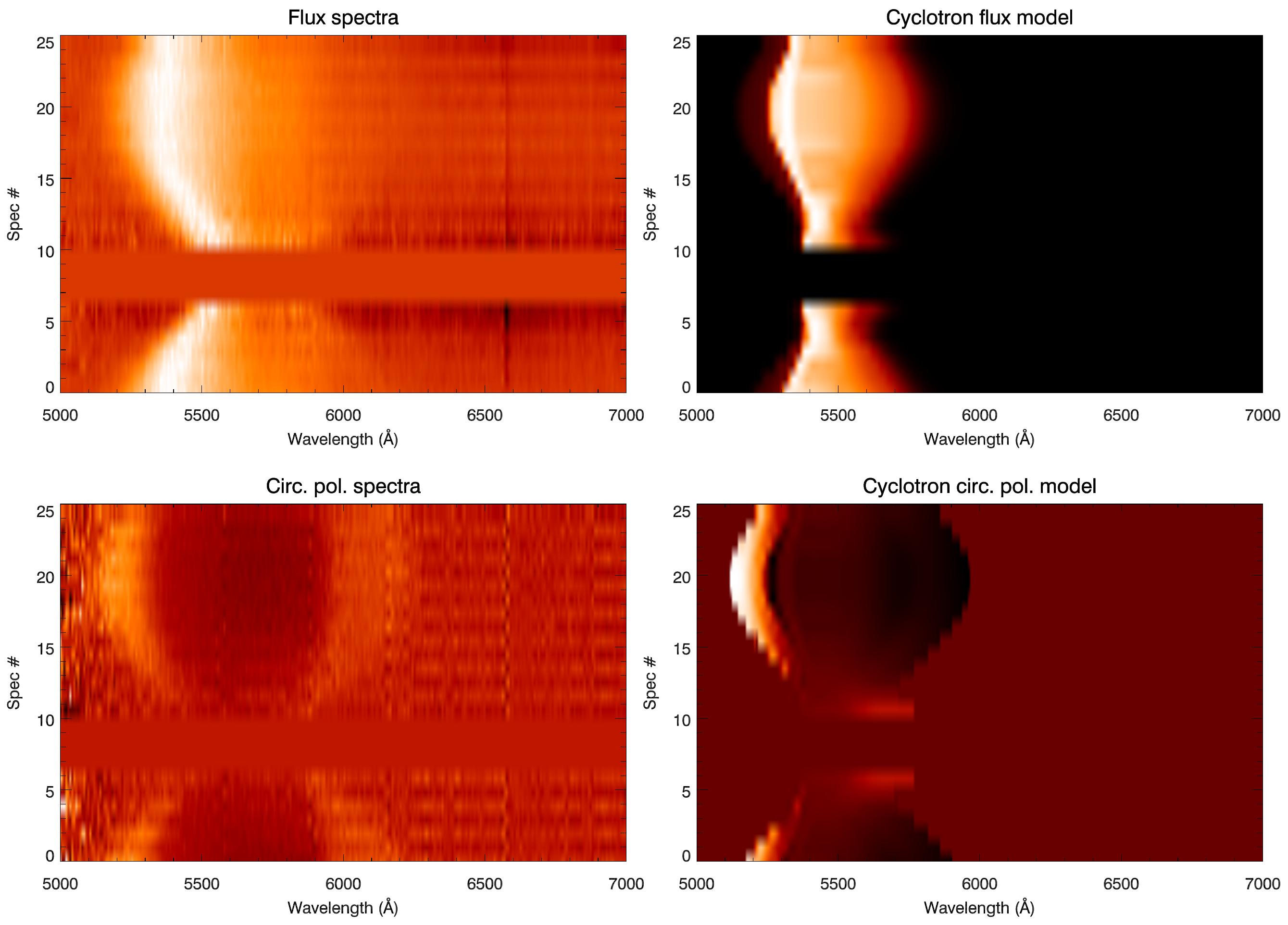}    
\vspace{-5mm}
  \caption{Circular spectropolarimetry of SDSS J0750+4943. The four top panels show the flux and circular polarisation of spectrum \#1 (left) and \#8 (right) highlighting the orbital phases where the cyclotron emission regions are
  visible and hidden behind the WD. The bottom four panels show the trailed cyclotron emission spectra and its circular polarisation (top two), together with our cyclotron model fits (bottom two). Note that there is a 0.5 orbital period gap between spectra \# 7 and 11. The flux and polarisation within this gap also mark the zero levels accordingly.}
    \label{figure4}
    \end{center}
\end{figure*}

SDSS J075015.11+494333.2 has an orbital period of 4.2\,h \citep{Drake2014} and was identified as a pre-polar candidate from its unusual light curve \citep{Parsons2015}. Optical spectra revealed strong cyclotron features making the source a very likely high field polar, with a magnetic field of either 99 or 196\,MG \citep{2021MNRAS.502.4305P}. We obtained 4.4\,h
of circular spectropolarimetry using a grism \#5 with ALFOSC on 4th Mar 2021. The polarimetric time resolution was 600 s resulting in 26 spectra over the orbital period (Fig. \ref{figure4}). The spectra cover a wavelength range of 5000-9500\AA. We detect a single cyclotron hump in this range, centered at around 5600-5700\AA. There is no indication of circular polarisation towards the red end of the spectrum. This implies that the detected cyclotron hump must correspond to the 2nd harmonic of the cyclotron base frequency with a magnetic field of 94-96\,MG, consistent with the  VLT/X-shooter results of \citet{2021MNRAS.502.4305P}. 

The circular polarisation behaviour of the cyclotron hump is intriguing in a similar way as in ZTF J0146+4916 i.e. the polarity shows sign reversal at both blue and red ends of the cyclotron hump. Again, like in ZTF J0146+4914, both the positive and negative poles seem to show the same orbital behaviour. Thus the same conclusions regarding the accretion geometry/physics apply here. As the observations of SDSS J0750+4943 are of much higher quality than the ZTF J0146+4914 data, we have attempted modelling of the spectropolarimetry using a simple model, consisting of two cyclotron emission regions on the WD surface. Both of these are assumed to have a ``slab" like geometry with radial magnetic field lines. The cyclotron emission spectrum, dictated by the viewing angle, surface magnetic field strength, the constant plasma temperature and a constant dimensionless plasma parameter $\Lambda$, is computed using the method (and code) described in \citet{1992A&A...256..498W},\citet{1993A&A...280..169W} and \citet{1996A&A...310..526R}.

In order to fit the spectropolarimetry, we have first taken advantage of the fact that the
cyclotron emission is not visible at all orbital phases. We can thus subtract away the contribution from the companion star from all the spectra showing evidence for cyclotron 
emission. We have then fitted simultaneously 24 flux spectra and the corresponding circular polarisation spectra from different orbital phases. The model includes two
point-like cyclotron regions that are free to move on the WD surface. In addition to their locations, the free parameters for the two cyclotron regions (separately) include: The magnetic field strength, shock
temperature, dimensionless plasma parameter $\Lambda$ and relative brightness of the regions. We attempted fitting with both: i) different values of fixed inclination and ii) the inclination as an additional free parameter. The fitting itself was carried out using a Differential Evolution (DE) algorithm \citep{SP1997}. The results are shown in Fig \ref{figure4}. Whilst we are able to obtain reasonable fits to the flux spectrum, the fits to the circular polarisation spectrum are not good. Furthermore, the resulting models place both of the cyclotron emission regions in the `southern' hemisphere of the WD (colatitudes depending on the inclination), separated only by a couple of degrees. Yet these two regions are required to have different magnetic polarity. In fact, the flux spectrum itself could be reasonably well fitted with just a single simple cyclotron region, if the polarisation would be ignored. Clearly, our simple model is not likely to represent the true picture and more detailed modelling would be required, perhaps employing a multipole magnetic field for the WD.

\subsection{{\it SDSS J085336.03+072033.5}}

\begin{figure}
  \begin{center}
%  \vspace{}
  \includegraphics[width=0.5\textwidth]{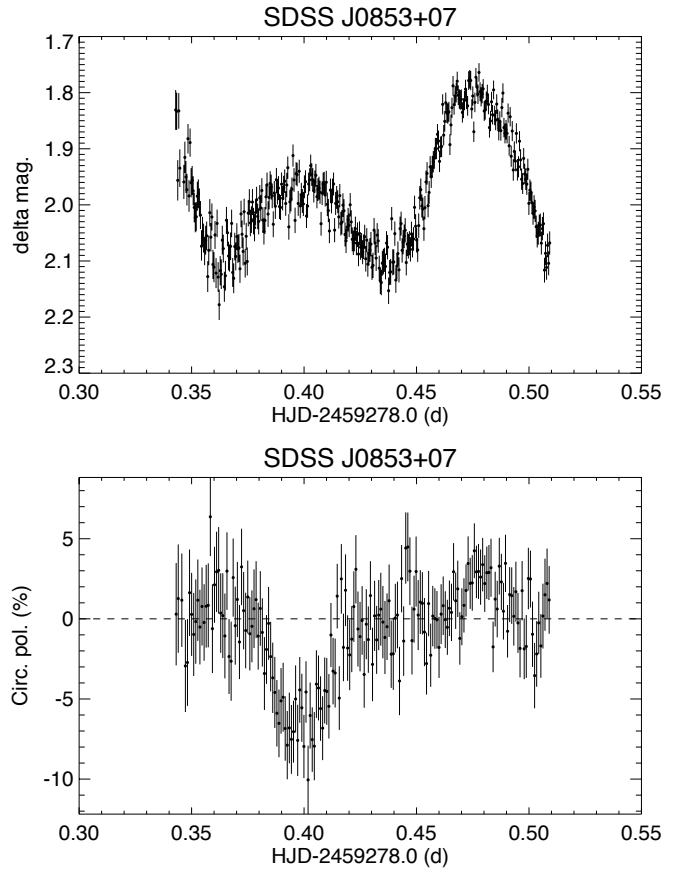}    
\vspace{0mm}
  \caption{Circular $R$ band photopolarimetry of SDSS J0853+0720. Top panel: the differential magnitude compared to a nearby comparison star. Lower panel: the circular polarised light curve.}
    \label{figure5}
    \end{center}
\end{figure}

SDSS J0853+0720 shows a 3.6\,h period \citep{Nebot2011}, with X-Shooter spectra revealing cyclotron features suggesting a magnetic WD with a field strength of 84\,MG \citep{2021MNRAS.502.4305P}. Our $R$ band photopolarimetric time series spans over a single 4\,h run (Fig \ref{figure5}). The light curve shows a clear overlap with repeating behaviour in the end. It is characterised by two maxima of unequal brightness. The lower maximum coincides with a clear maximum in negative circular polarisation, which reaches about $\sim$8\%. The circular polarisation is reversed during the higher maximum but, somewhat unexpectedly, the circular polarisation only reaches about +3\,\% and lasts for much less than the duration of the hump in the light curve. Nevertheless, we suggest this system is accreting on two poles, the visibility of which alternates over the orbital period. Finally, the $R$ band's effective wavelength of 6400\,\AA\, coincides with the centre wavelength of
the second harmonic cyclotron hump from the 84\,MG field.  

\begin{figure}
  \begin{center}
  \hspace{0mm}
  \includegraphics[width=0.49\textwidth]{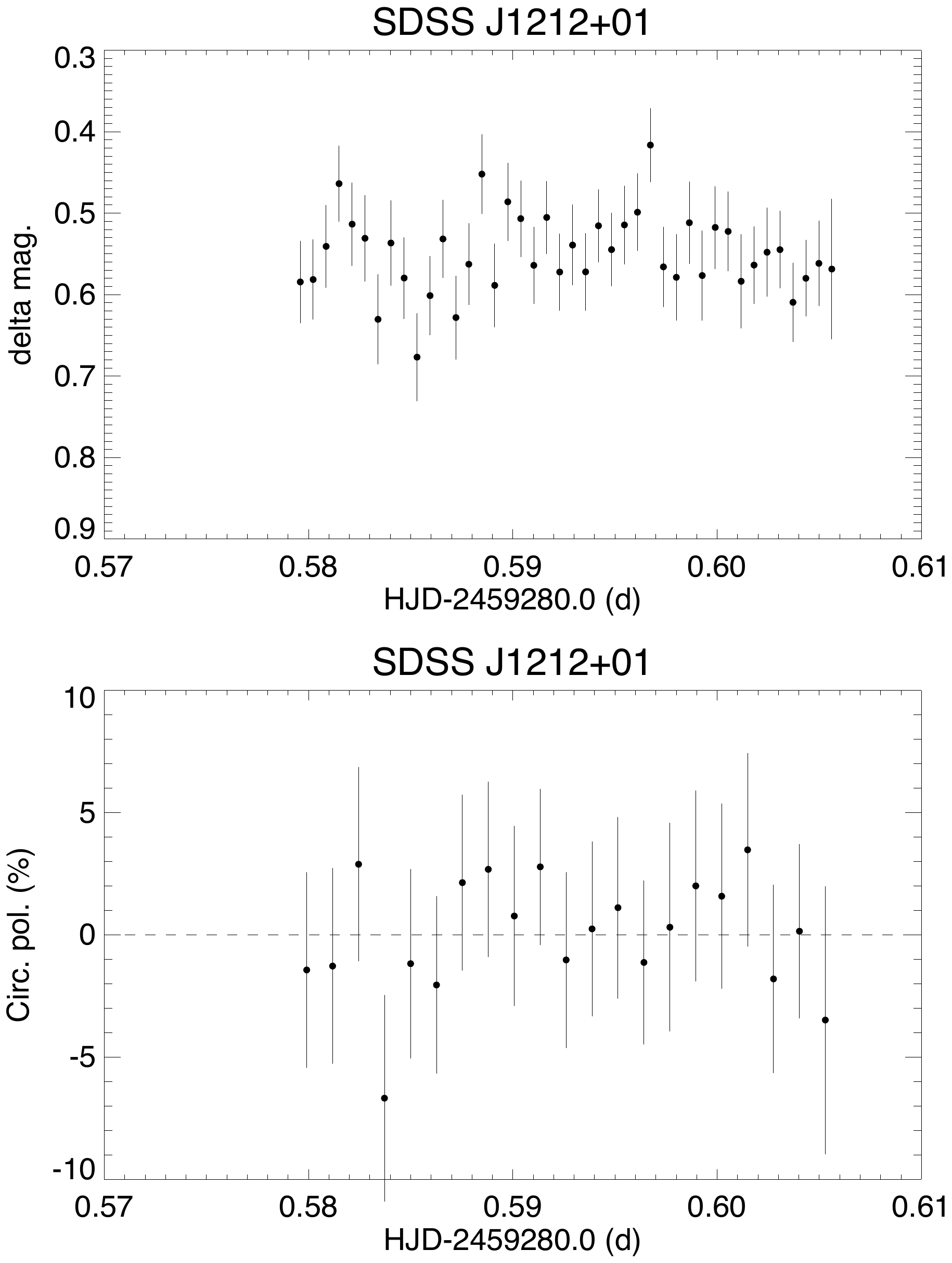}    
\vspace{-5mm}
  \caption{Circular photopolarimetry of SDSS J1212+0136. Top panel: the differential magnitude compared to a nearby comparison star. Lower panel: the circular polarised light curve.}
    \label{figure6}
    \end{center}
\end{figure}

\begin{figure*}
  \begin{center}
  \hspace{0mm}
  \includegraphics[width=0.95\textwidth]{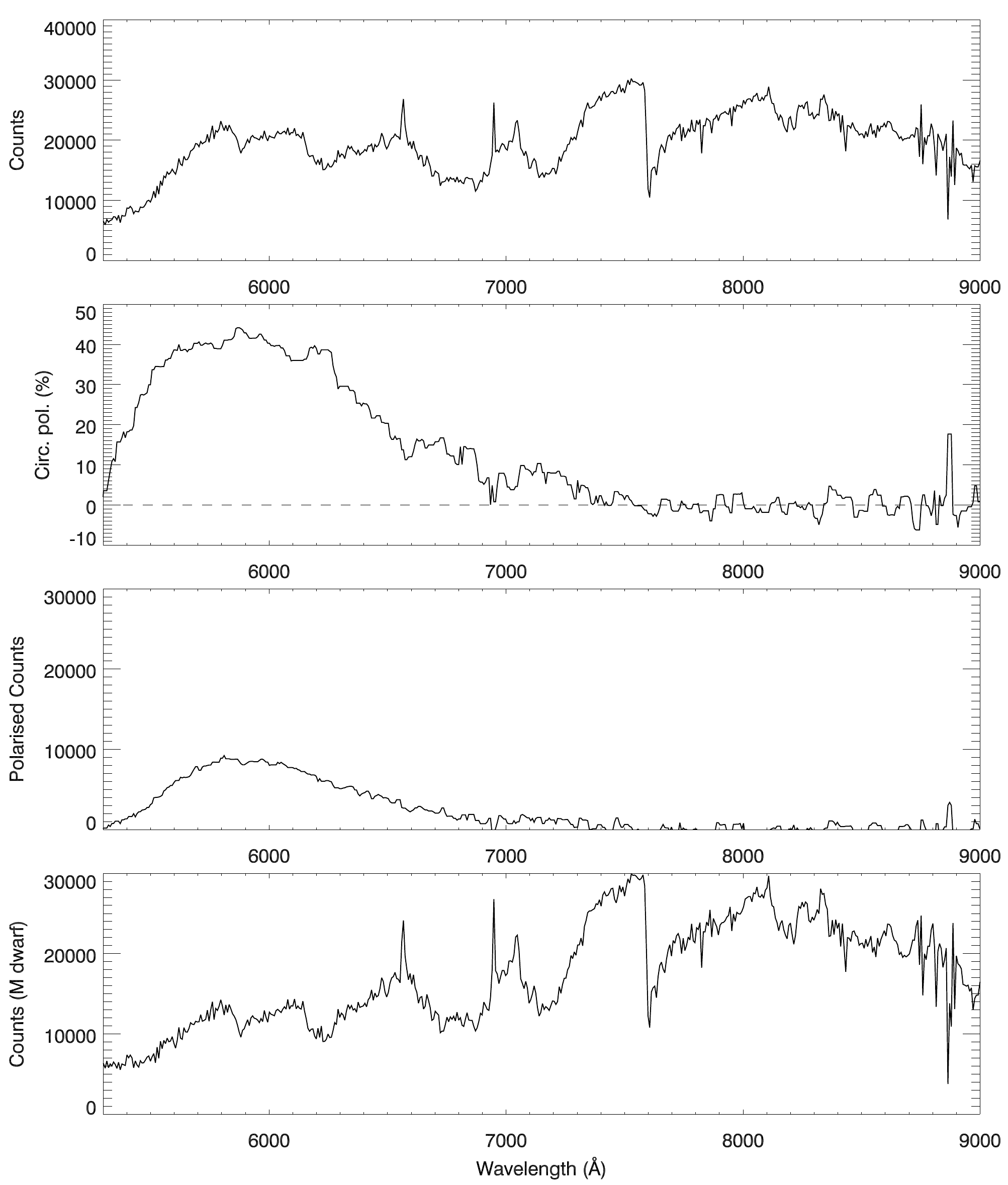}    
\vspace{0mm}
  \caption{Circular spectropolarimetry of SDSS J2229+1853, The panels (from top) show the count spectrum, the circular polarisation spectrum, the polarised counts and the unpolarised count spectrum, associated mostly with the M dwarf. A very prominent cyclotron hump centered at around 5800-6000 \AA\ is present.}
    \label{figure8}
    \end{center}
\end{figure*}

\begin{figure}
  \begin{center}
  \hspace{-5mm}
  \includegraphics[width=0.5\textwidth]{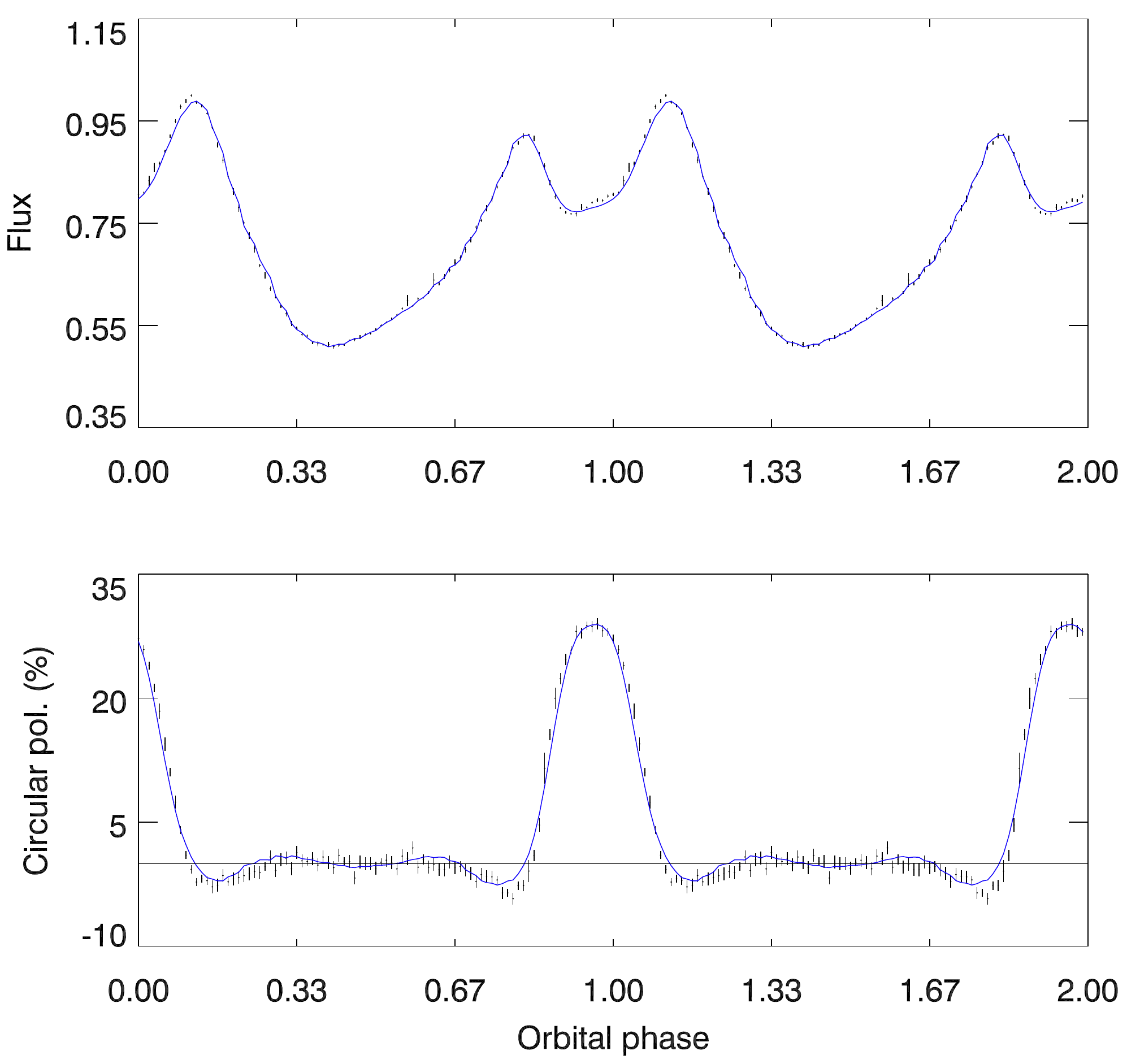}   
  \hspace{0mm}
  \includegraphics[width=0.48\textwidth]{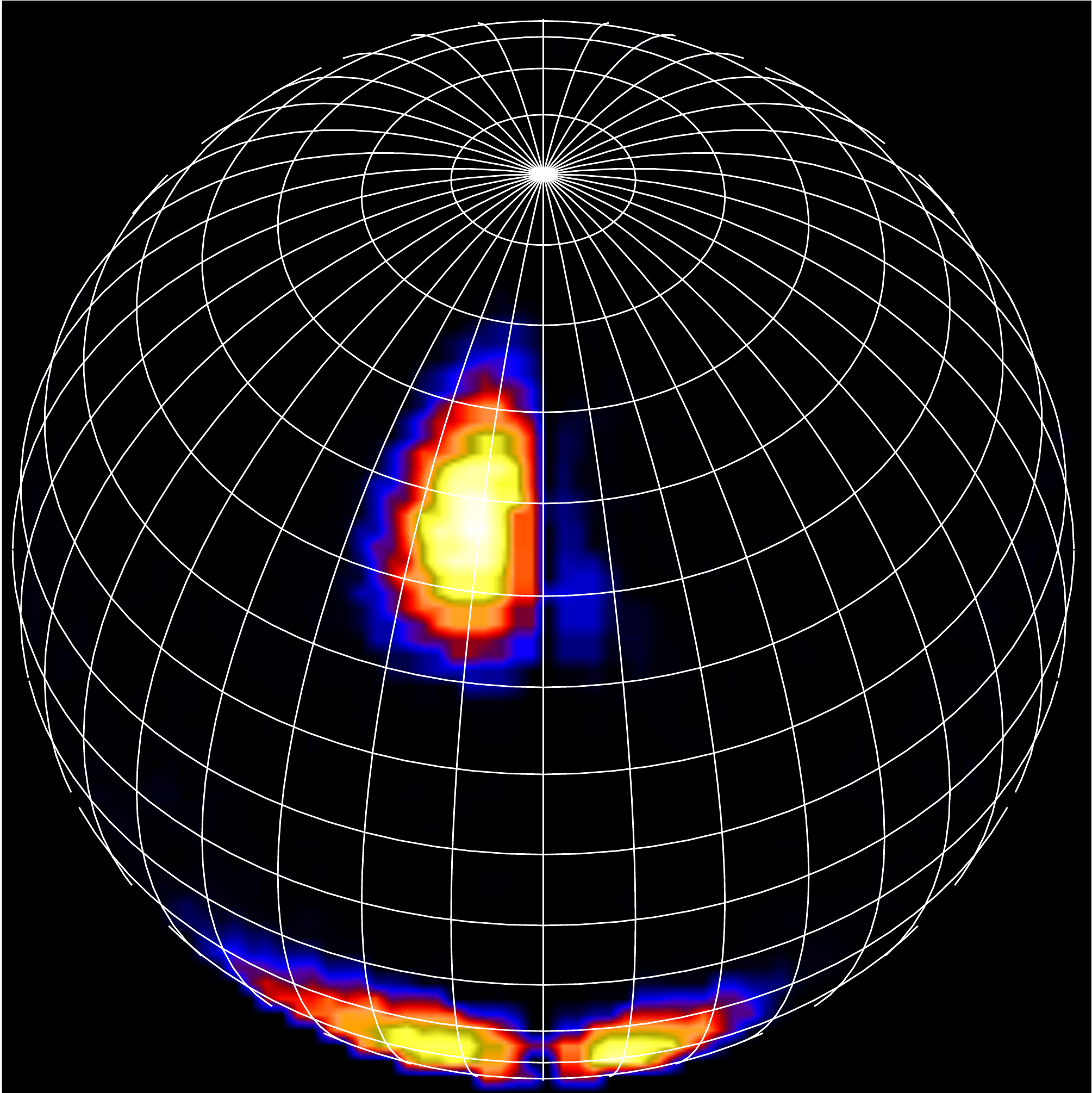}    
%\vspace{-35mm}
  \caption{Top panels: Circular photopolarimetry of SDSS J2229+1853 with a cyclotron emission model fit. Lower panel: the implied cyclotron emission map of the WD surface. The viewing angle to the WD is shown as at $i$=45, phase=0.0}
    \label{figure9}
    \end{center}
\end{figure}

\subsection{{\it SDSS J121209.31+013627.7}}

This is a very likely brown dwarf + magnetic WD binary with an orbital
and WD spin period of 88 min \citep{2003ApJ...595.1101S,2005ApJ...630L.173S,Burleigh2006}. 
\citet{2017A&A...598L...6S} confirmed it as weakly accreting polar-like object through time-resolved X-ray observations with XMM-Newton.
A WD magnetic field of 7\,MG (mean) and 13\,MG (dipole) were reported by \citet{2003ApJ...595.1101S} and supported based on NIR spectroscopy \citep{Farihi2008}. Given the low magnetic field estimate, we obtained $i$ band data of this system (Fig.\ref{figure6}). However, the observations were cut short due to bad weather and the data cover 
only 52 min (i.e. 0.59 in orbital phase). We do not see any evidence for circular polarisation, but it is possible the polarisation signal was missed due to the
limited phase coverage. Furthermore, the observations suffered from thin cirrus.

\subsection{{\it SDSS J222918.95+185340.2}}

SDSS J222918.95+185340.2 was identified as a potential pre-polar candidate from its unusual CRTS light curve \citep{Drake2014}. X-shooter spectra covering 2/3 of the orbital cycle reveal at least 6 cyclotron lines appearing at different phases, suggesting a complicated field structure \citep{2021MNRAS.502.4305P}. Optical light is very dominated by the M3 companion star in the very red end of the spectrum. The binary star's orbital period is 4.5\,h and the system is extremely close to Roche-lobe filling (the fit is actually consistent with Roche-lobe filling) but beyond the cyclotron lines there are no other obvious signs of accretion
\citep{2021MNRAS.502.4305P}.

We obtained 4 circularly polarised spectra of SDSS J2229. The summed up flux and circular polarisation spectra are shown in Fig \ref{figure8}. We detect a clear cyclotron hump centered around 6000\,\AA. 
There is no evidence in circular polarisation for any other cyclotron hump up to 10000\,\AA. This strongly suggests that the 6000\,\AA\ hump is likely the second harmonic component, arising from the 85-90\,MG WD magnetic field, consistent with the X-shooter 
data \citep{2021MNRAS.502.4305P}.

The $R$ band circular time resolved photopolarimetry is shown in Fig \ref{figure9}. The circular polarisation reaches 30\% at its peak, compatible with the spectropolarimetric results (Fig \ref{figure8}). Perhaps the most 
striking feature of the photopolarimetry is the total absence of any flickering in the data. This is likely due to wind accretion, as there is no evidence for any accretion stream or Roche lobe 
overflow in the system (i.e. no emission lines or signs for accretion stream continuum emission). It is feasible that accreted stellar wind lacks the clumpiness of accretion 
flow resulting from Roche lobe overflow. 

We have used the resulting high signal-to-noise flux and polarisation light curves to map the WD surface in cyclotron radiation. This approach is very similar
to the 'Stokes Imaging' \citep{Potter1998}, although we are lacking the linear polarisation information. The approach is also related to the one used for TESS CD Ind data \citep{Hakala2019}, although the TESS data obviously lacked any polarisation information. 

In brief, we have utilised the cyclotron emission models of \citet{WM1985} to model the dependence of the cyclotron flux and circular polarisation as a function of angle between the line of sight and 
the magnetic field line. We have used a 10\,keV, 2nd harmonic model with $\Lambda=10^5$ for this.
Since we do not have information on the magnetic field geometry of the WD, we have assumed that 
the cyclotron emission originates near the magnetic poles, where the magnetic field lines approximately align with the surface normal. We have also assumed that the cyclotron emission is
not confined to the surface of the WD, but the emission regions can have height up to 6\% of the WD 
radius (corresponding to observing the shock front up to 20$^{\circ}$ beyond the limb crossing). This is directly indicated by the data, as circular polarisation sign reversals are seen before and after the
limb crossings (i.e. the cyclotron emission region is seen from below). As there is no evidence for
an obvious second pole (with negative circular polarisation), we have assumed that all the field lines, where the emission is detected, flow radially outwards. However, we have also experimented with sign reversal of magnetic field for different hemispheres, but that failed to produce acceptable fits. For the actual fitting, we used a Differential Evolution (DE, \citealt{SP1997}) algorithm to
optimize a smoothest possible cyclotron emission map over the WD surface. This was done using a grid
with 4$^{\circ}$ resolution (see \citealt{Hakala2019} for more details). The fitting was carried out with fixed inclination angle. Several inclination angles were tried, together with varying amount of unpolarised additional flux. 

The best results were obtained using $i=45^{\circ}$ and no additional unpolarised light. This strongly suggests there is no contribution from an accretion stream which, if present, typically accounts for large fraction of optical emission, as evidenced by eclipse profiles \citep{Hakala1995} and Doppler mapping \citep{1999MNRAS.304..145H}. The resulting cyclotron map is shown in the bottom panel of Fig \ref{figure9}. The cyclotron emission appears to be dominated by a main accretion region in the `Northern' hemisphere at a latitude of 40-50$^{\circ}$ and pointing
almost towards the secondary star. There are two additional weaker regions roughly at the same WD longitude, but at latitude -20$^{\circ}$. These secondary regions are thus separated by $\sim$60-70$^{\circ}$ from the main region. Given the small gap 
between these regions, the separation into two close regions is probably not significant, but likely arises from the regularisation. Remarkably, the locations
of the main cyclotron regions Fig \ref{figure9} (at different WD hemispheres, both facing the donor star) are almost identical to the ones measured in another similar system SDSS J030308.35+005444.1 \citep{Parsons2013} using eclipse data, even if the magnetic field here is an order of magnitude greater. However, we should be careful in putting too much weight on interpreting these results though, as the true magnetic field configuration is not known nor implemented in our modelling, which assumes radial field lines at cyclotron emission sites.

\begin{table*}
  \begin{tabular}{llllll}
    \hline
Source & P$_{Orb}$ & Details & Peak circ.pol. & Inferred/known {\bf B}-field & Remarks \\
    \hline
2MASS J01294349+671530$^1$ & 7.15\,h & K7V+WD, ecl. & 4\% & 81-85\,MG (or 53-57\,MG) & 53-57\,MG field cannot be ruled out. \\

ZTF J014635.73+491443.1$^4$ & 2.057\,h & BD?+WD & $\sim$40\% & 89\,MG$^4$ & Distance 56.3 $\pm$ 0.3\,pc \\

SDSS J030856.55$-$005450.7$^3$ & 4.5\,h & M4+WD, ecl. & No detection & N/A &  \\

SDSS J075015.11+494333.2$^2$ & 4.2\,h & M2.5+WD & $\sim$20\% & 94-96\,MG$^2$ & Complex magnetic field geometry implied.\\  	

SDSS J085336.03+072033.5$^2$ & 3.6\,h & M4+WD & 8\% & 84\,MG$^2$ & two pole accretion.	\\

SDSS J121209.31+013627.7$^5$ & 1.47\,h & BD+WD & No detection & 13\,MG$^5$ & Observations covered only 0.59 in phase \\

SDSS J222918.95+185340.2$^2$ & 4.5\,h & M3+WD & 45\% & 85-90\,MG & Cyclotron mapping reveals two-pole accretion.\\ 

\hline
  \end{tabular}
  \caption{The summary of system parameters and main results. .($^1$\citet{Krush2020},$^2$\citet{2021MNRAS.502.4305P},$^3$\citet{Becker2011},$^4$\citet{Guidry2021},$^5$\citet{2003ApJ...595.1101S}}
  \label{table1}
\end{table*}

\begin{figure}
  \begin{center}
  \hspace{0mm}
  \includegraphics[width=0.53\textwidth,angle=0]{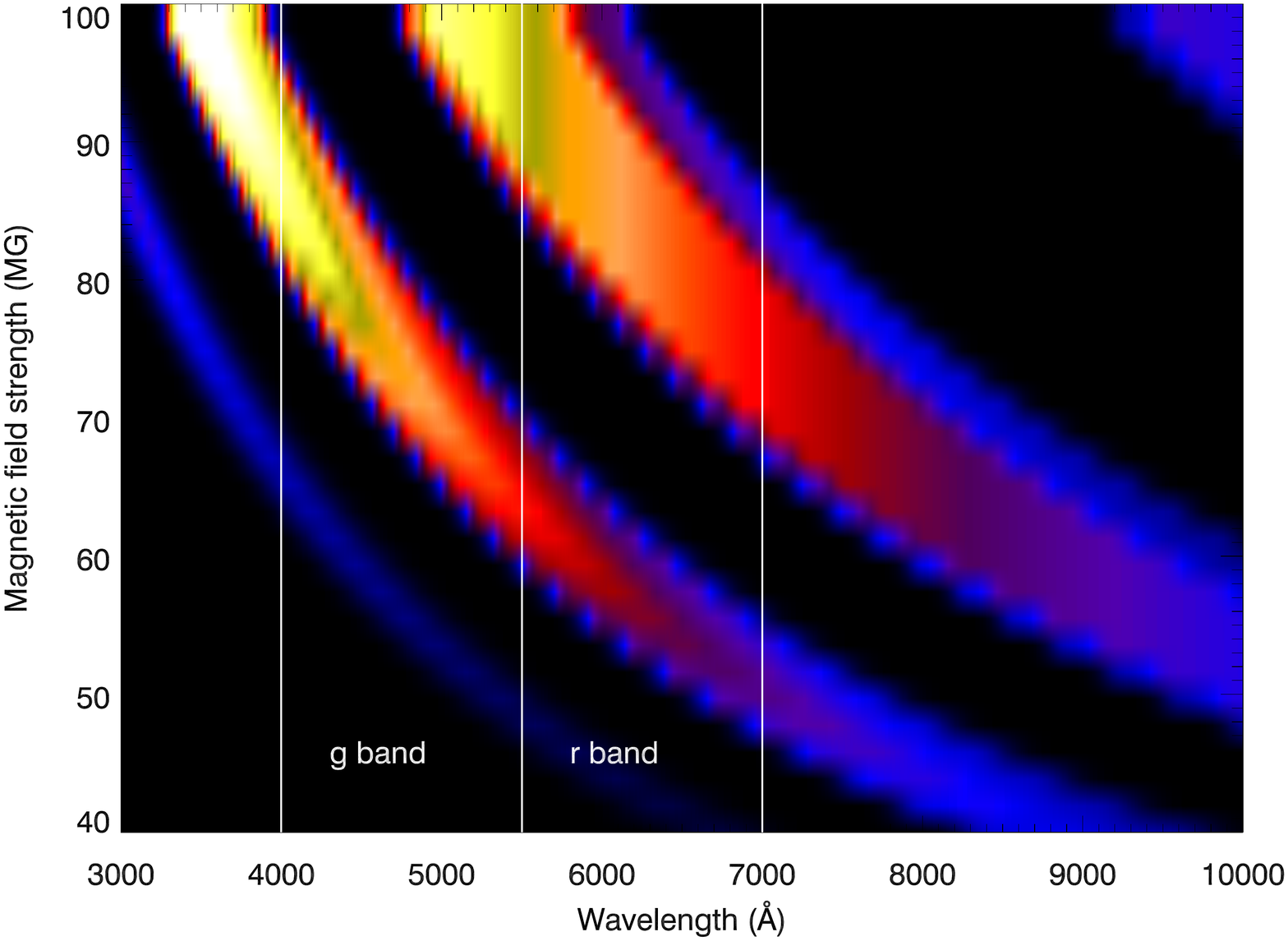}   
%\vspace{-35mm}
  \caption{The cyclotron emission spectra from constant low temperature (1keV) and ($log \Lambda = 5$)  ``slab" models as a function of the magnetic field strength. The locations of g and r passbands are shown with vertical dashed lines.}
    \label{figure10}
    \end{center}
\end{figure}

\section{Conclusions}

We have obtained the first optical circular imaging and spectropolarimetry of seven pre-polar (or low accretion rate polar) candidates. The sources were chosen based on either existing optical spectroscopy or optical light curve shapes suggesting cyclotron emission in the systems. None of the chosen systems show strong emission lines or any other evidence of accretion taking place and light curve modelling of three of these systems \citep{2021MNRAS.502.4305P} indicate that the secondary stars do not fill their Roche lobes. We find that in 5 out of 7 systems we detect variable optical circular polarisation, confirming the cyclotron emission and thus the presence of the magnetic WD. The summary of our polarimetric results can be found in Table 2.

Intriguingly, all of the 5 systems exhibiting circular polarisation show WD magnetic fields in excess of 80 MG.  
This is considerably higher than is typically found in Roche lobe overflow accreting polars 
\citep{2015SSRv..191..111F}. In fact \citet{2015SSRv..191..111F} only list four polars (or pre-polars) out of 72 that show WD field strength in excess of 80 MG. This suggests that maybe the pre-polar stage systems with the highest field WDs are more easily detected than the lower
field pre-polars. This could be due to the more effective channeling of the stellar wind from the secondary star on to the WD. The second option is that the WDs in pre-polars indeed are intrinsically divergent and have experienced a different formation and/or accretion history than the WDs in polars. Finally, the most likely case is perhaps that this is due to a selection effect. \citet{2009A&A...500..867S} discuss the similar difference in WD properties regarding their sample or 9 LARPs that showed higher then average magnetic fields, cool WDs and longer orbital periods than the known polars. They conclude that selection effects are to likely play a significant role, as the accreting plasma in pre-polars is tenuous and cooler than in polars, that sustain much higher accretion rates. This leads to the cyclotron emission peaking in the lower harmonics and only weak emission from harmonics higher than 3rd is observed. 

We demonstrate the effect
in Fig \ref{figure10}, where we have plotted the cyclotron spectra from constant temperature models (log $\Lambda$=5 and T$_{eff}$ = 1 keV, viewing angle = 45$^\circ$) with magnetic field strength ranging from 40 to 100 MG. The two strong cyclotron humps correspond to the 2nd and 3rd
harmonics. We also show the Sloan g and r passbands separated by vertical dashed lines. It is clear
that the systems containing more magnetic WDs are much more easily picked up based on say Sloan passband photometry and/or optical spectroscopy. The same bias is not equally obvious for the 
polars, where the shock region near the WD surface is much hotter and denser, leading to stronger
cyclotron emission at harmonics higher than the 3rd. Thus it is likely that the range of WD magnetic fields in our sample of systems is at least partly due to the fact that our sample is selected based on possible cyclotron features in the spectra or light curves of the candidates.

It is worth mentioning that, even if the fundamental frequency of cyclotron emission is generally thought to be optically thick and non-polarised in polars, this might not
be the case in wind-accreting systems, where the accretion rates are several magnitudes lower \citet{2002ASPC..261..102S}. Thus it is feasible, based on our cyclotron model computations, that the emission at the fundamental frequency becomes optically thin and exhibits significant circular polarisation if the plasma density becomes low enough (log$\Lambda \lesssim$ 2.5 in our modelling). 
The implication is that the magnetic field estimates based on seeing a single cyclotron 
hump, with no polarisation towards the red end of the optical spectrum, could be too low by a factor of two, in case the observed hump would in fact represent emission at the fundamental frequency. This applies to sources 2MASS\,J0129+6715, SDSS J0750+4943 and  
SDSS J2229+1853.

We have attempted to model the cyclotron emission in detail in two of the systems. The spectropolarimetric modelling of SDSS J0750+4943 failed to produce satisfactory results suggesting a complex WD field geometry. However, the photopolarimetric modelling of SDSS J2229+1853 was successful and implied two elongated cyclotron emission region located roughly at the same longitudes, but with 70-80$^{\circ}$ separation in latitude. The light
curve is entirely dominated by the cyclotron emission, with no additional emission whatsoever. It is worth noting that \citet{2021MNRAS.502.4305P} measured the accretion rate for SDSS J2229+1853 (based on cyclotron flux). They found $6.5\times10^{-13}$M$_{\odot}$ yr$^{-1}$, which provides the lower limit for the stellar wind from the M3V companion star. This value is very high ($30 \dot M_{\odot}$), but such winds have been measured \citep{2021ApJ...915...37W}.
However, the likely implication is that almost all of the stellar wind is accreted and, at least
in its current state, the magnetic braking should be very inefficient.
The total absence of any flickering on top of the cyclotron emission is a strong argument against the typical accretion scheme in polars and perhaps serves as an indication of accretion via stellar wind. 

Apart from 2MASS 012943+6715, and ZTF J0146+4914, all the other three systems containing a highly magnetic WD are compatible with the pre-polar scheme as outlined in \citet{2021NatAs...5..648S}. The suggested asynchronism in 2MASS 012943+6715 needs to be verified by further polarimetric observations.  
Finally, we detect a very likely brown dwarf companion in the period bouncer
system ZTF J0146+4914. We estimate that the companion brown dwarf is possibly of late T spectral type, slightly heated by the WD, to match the WISE observations. 

\section*{Acknowledgements}

Based on observations made with the Nordic Optical Telescope, owned in collaboration by the University of Turku and Aarhus University, and operated jointly by Aarhus University, the University of Turku and the University of Oslo, representing Denmark, Finland and Norway, the University of Iceland and Stockholm University at the Observatorio del Roque de los Muchachos, La Palma, Spain, of the Instituto de Astrofisica de Canarias.
The data presented here were obtained with ALFOSC, which is provided by the Instituto de Astrofisica de Andalucia (IAA) under a joint agreement with the University of Copenhagen and NOT. We would like to thank the NOT staff for carrying out the observations. Armagh Observatory and Planetarium is core funded by the Northern Ireland Executive through
the Dept. for Communities. SGP acknowledges the support of a Science and Technology Facilities Council (STFC) Ernest Rutherford Fellowship.213056.71+442046.5. TRM and BTG were supported by grant ST/T000406/1 from the Science and Technology Facilities Council (STFC). 
This project has received funding from the European Research Council (ERC) under the European Union’s Horizon 2020 research and innovation programme (Grant agreement No. 101020057). 
This research has made use of the VizieR catalogue access tool, CDS, Strasbourg, France (DOI : 10.26093/cds/vizier). The original description of the VizieR service was published in 2000, A\&AS 143, 23. For the purpose of open access, the author has applied a Creative Commons Attribution (CC BY) licence to any Author Accepted Manuscript version arising. Finally, we wish to thank the referee, Prof. John Landstreet, for detailed remarks that certainly improved the paper considerably.

\section*{Data Availability}

The data will be shared on reasonable request to the corresponding author.

\vspace{4mm}

\bibliographystyle{mnras}

% Don't change these lines
\bsp	% typesetting comment
\label{lastpage}

\end{document}